\documentclass[english,twoside,a4paper,10pt]{article}

\usepackage[latin1]{inputenc}
\usepackage[T1]{fontenc}
\usepackage{amsmath}
\usepackage{amsfonts}
\usepackage{graphicx}
\usepackage{subfigure}
\usepackage{a4wide}
\usepackage{amssymb}
\usepackage{fancyhdr}
\usepackage{mathrsfs}
\usepackage[toc,page]{appendix}

\begin{document}

\title{\bf $F(R)$-gravity and inflation}
\author{ 
Lorenzo Sebastiani\footnote{E-mail address: l.sebastiani@science.unitn.it
},\,\,\,
Ratbay Myrzakulov\footnote{Email: rmyrzakulov@gmail.com}\\
\\
\begin{small}
Department of General \& Theoretical Physics and Eurasian Center for
\end{small}\\
\begin{small} 
Theoretical Physics, Eurasian National University, Astana 010008, Kazakhstan
\end{small}\\
}

\date{}

\maketitle

\begin{abstract}
In this short review, we revisit inflation in $F(R)$-gravity. We find several $F(R)$-models for viable inflation 
by applying some reconstruction techniques. A special attention is payed in the reproduction of the last Planck satellite data. The possible generalizations of Starobinsky-like inflation are found and discussed. The early-time acceleration is analyzed in a higher-derivative quantum gravitational model which mainly reduces to $F(R)$-gravity.
\end{abstract}



\tableofcontents
\section{Introduction}

Today it is well accepted the idea according to which the universe underwent a period of strong accelerated expansion, namely the inflation, after the Big Bang. Inflation was introduced several years ago by Guth~\cite{Guth} and Sato~\cite{Sato} in the attempt to solve some problems related to the initial conditions of Friedmann universe. In particular, the cosmic microwave background (CMB) shows that the observable universe is thermalized with high accuracy, implying that, back into the past, it was causally connected: an early-time acceleration can explain this fact in agreement with the following radiation/matter dominated expansion (see Refs.~\cite{Linde, revinflazione} for reviews).

The early-time acceleration is clearly supported by a repulsive force and is the result of a quasi-de Sitter expansion  at high curvature  (the ``false vacuum''). In order to recover the radiation/matter dominated universe inflation must quickly finish, and in order to reproduce the inhomogeneities of our universe perturbations at the end of inflation must be in agreement with the anisotropy spectrum from cosmological observations~\cite{WMAP, Planck, Planck2}.

The arena of inflationary models is quite large. For example, in scalar field theories, a scalar field (the inflaton)
subjected to some suitable potential drives the de Sitter expansion when its magnitude is negative and very large: this is the chaotic scenario~\cite{ buca1, buca2, chaotic, buca3, buca4}. At the end of inflation, the field falls in a minimum of the potential and starts to oscillate such that the reheating processes for particle production take place.

One may expect that inflation comes from a fundamental theory of gravity and is the effect of some quantum correction of Einstein's theory at high energy. In modified theories of gravity~\cite{reviewmod, reviewmod2, Caprev, myrev} higher-derivative terms of the curvature invariants are considered in the gravitational action. The study of modified gravity is quite interesting, even because, due to the large choice of the models, one can recover the phenomenology of many different theories. For example, $F(R)$-gravity, where the modification of the action of General Relativity is in terms of the Ricci scalar only, has a corresponding representation in the (one single) scalar field framework. An example is the so called ``Starobinsky model'' for inflation~\cite{Staro} with the account of a correction quadratic in the Ricci scalar in the modified gravity framework, and of an exponential potential in the scalar field framework.

In this short review, we would like to revisit inflation in $F(R)$-gravity. Viable models in agreement with last cosmological data are investigated in the ``Jordan-frame'' of modified gravity and in the corresponding ``Einstein-frame'' of scalar field representation. Inflation behind the Starobinsky model is analyzed.
In the last part of the work, we will study the early-time acceleration in a higher-derivative gravity theory with the account of quantum effects. 
The aim of this work is to recall the results obtained in $F(R)$-gravity for inflation by starting from the cosmological data or by deforming the models that already reproduce such data: as we mentioned before, despite to the fact that accelerated expansion can be supported by a large class of models, the observations, and, in the specific, the match between viable spectral index and the tensor-to-scalar ratio, play a fundamental role in the choice of the theory, leading to a restrict class of Lagrangians for realistic inflation.
This work is based on Refs.~\cite{uno, due, tre}. 

The paper is organized in the following way. In Section {\bf 2}, we remind some basic features of inflation. In Section {\bf 3}, we present the transformation  formulas to pass from Jordan- to Einstein-framework in $F(R)$-gravity. Moreover, we furnish an unified description of inflation in the both representations. Section {\bf 4} is devoted to the reconstruction of viable inflation by starting from the spectral index and the tensor-to-scalar ratio inferred from the last Planck satellite data. In Section {\bf 5}, we reconstruct inflation in $F(R)$-gravity by starting from viable inflation in scalar field theories. In Section {\bf 6}, we analyze the early-time acceleration from a renormalization group improved effective Lagrangian which can be reduced in the form of $F(R)$-gravity. Conclusions and final remarks are given in Section {\bf 7}.

We use units of $k_{\mathrm{B}} = c = \hbar = 1$ and denote the
gravitational constant, $G_N$, by $\kappa^2\equiv 8 \pi G_{N}$, such that
$G_{N}^{-1/2} =M_{\mathrm{Pl}}$, $M_{\mathrm{Pl}} =1.2 \times 10^{19}$ GeV being the Planck mass.

\section{Some facts about the early-time acceleration}

Thanks to the cosmic acceleration, small initial velocities within a causally connected patch become very large. In this way, the inflationary paradigm, where the universe undergoes an early-time accelerated expansion, is able to explain the thermalization of our obervable universe, and solves some problems related to the initial conditions of Friedmann cosmology. Inflation takes place after the Big Bang, during a time corresponding to the Planck epoch 
 ($t\sim10^{-35}-10^{-36}$ sec), and can be described by a (quasi) de Sitter expansion where the Ricci scalar is almost a constant and is near to the Planck scale. 

A useful parameter to describe inflation is
the $e$-folds number left to the end of inflation,
\begin{equation}
N=\ln\left[\frac{a(t_\text{f})}{a(t)}\right]\,,\label{efolds}
\end{equation}
where $a(t_\text{f})$ is the scale factor at the end of inflation with $t_\text{f}$ the related time.
Thus, the total amount of inflation is given by
\begin{equation}
\mathcal N\equiv N |_{a(t_\text{i})}
\label{Nfolds}
\end{equation}
where 
$a(t_\text{i})$ is the scale factor at the beginning of inflation with $t_\text{i}$ the corresponding time. In order to get the thermalization of our universe according with CMB data, one has to require
\begin{equation}
55<\mathcal N<65\,.\label{rangeN}
\end{equation} 
A model for inflation must also possess a valid mechanism to exit from accelerated phase and to lead to the cosmological perturbations at the origin of the anisotropy of the universe. In order to measure
such perturbations one calculates the spectral index $n_s$ and the tensor-to-scalar ratio $r$, whose values are well determined by the last Planck data as~\cite{Planck2}: 
\begin{equation}
n_s = 0.968 \pm 0.006\, (68\%\,\mathrm{CL})\,,\quad 
r < 0.11\, (95\%\,\mathrm{CL})\,.\label{viablespectral}
\end{equation}
These indexes have to been derived in different way depending on the theory under consideration, and in what follows we will furnish their correct expressions in modified gravity and in scalar field theories.

\section{$F(R)$-modified gravity and scalar field inflation}

Modifications to gravity are expected during inflation, where 
some corrections to Einstein's gravity may emerge at high curvature and support the early-time acceleration. The simplest class of modified gravity theory is given by $F(R)$-gravity, where a function of the Ricci scalar only replaces the Hilbert-Einstein term of General Relativity into the action. 

When we deal with $F(R)$-gravity, we can pass to its scalar-field representation to achieve some useful simplification: this is the case of $F(R)$-gravity for inflation, with a corresponding description in terms of chaotic inflation from scalar field, whose mechanism has well studied in literature.
The demonstration of the 
conformal equivalence between General Relativity plus scalar field contribute and $F(R)$-gravity was first done in 1988 by Barrow \&
Cotsakis~\cite{Spiros} and independently about the same time by 
Maeda~\cite{MaedaV}.

To
recast $F(R)$-gravity in the scalar field framework we must operate a conformal transformation on the metric. We remember that a general conformal transformation of the metric tensor $g_{\mu\nu}$ can be expressed as
\begin{equation}
\tilde g_{\mu\nu}=\Omega(t, {\bf x}) g_{\mu\nu}\,,\label{genconf}
\end{equation}
where $\Omega(t, {\bf x}) $ is a function of the space-time coordinates.

The action of $F(R)$-modified gravity in the so called ``Jordan Frame'' reads
\begin{equation}
I = \int_\mathcal{M} d^4 x \sqrt{-g} \left[ \frac{F(R)}{2\kappa^2}\right]\,,
\label{action}
\end{equation}
where $F(R)$ is a generic function of the Ricci scalar $R$, $g$ is the determinant of 
the metric tensor and $\mathcal M$ is the space-time manifold.
One can introduce the field $A$ as
\begin{equation}
I=\frac{1}{2\kappa^{2}}\int_{\mathcal{M}}\sqrt{-g}\left[
F_A(A) \, (R-A)+F(A)\right] d^{4}x\label{JordanFrame}\,.
\end{equation}
Here, we are using the following definition
\begin{equation}
F_{A}(A)=\frac{d F(A)}{d A}\,,
\end{equation}
such that, by making the variation of (\ref{JordanFrame})
with respect to $A$, we immediatly get $A=R$.
Thus, a scalar field $\phi$, which will be the inflaton in the scalar field representation and will encode the freedom degree from modified gravity, can be introduced as
\begin{equation}
\phi := -\sqrt{\frac{3}{2\kappa^2}}\ln [F_A(A)]\,.\label{sigma}
\end{equation}
Now, if we consider the conformal transformation of the metric (\ref{genconf}) with $\Omega(t,{\bf x})= \exp\left[-\sqrt{2\kappa^2/3}\phi\right]$,
\begin{equation}
\tilde g_{\mu\nu}=\mathrm{e}^{-\sqrt{2\kappa^2/3}\phi}g_{\mu\nu}\label{conforme}\,,
\end{equation}
we easily derive the `Einstein frame' action of the theory,
\begin{eqnarray}
I &=& \int_{\mathcal{M}} d^4 x \sqrt{-\tilde{g}} \left\{ 
\frac{\tilde{R}}{2\kappa^2} -
\frac{1}{2}\left(\frac{F_{AA}(A)}{F_A(A)}\right)^2
\tilde{g}^{\mu\nu}\partial_\mu A \partial_\nu A - 
\frac{1}{2\kappa^2}\left[\frac{A}{F_A(A)}
- \frac{F(A)}{F_A(A)^2}\right]\right\} \nonumber\\ 
&=&\int_{\mathcal{M}} d^4 x \sqrt{-\tilde{g}} \left( 
\frac{\tilde{R}}{2\kappa^2} -
\frac{1}{2}\tilde{g}^{\mu\nu}
\partial_\mu \phi \partial_\nu \phi - 
V(\phi)\right)\,,\label{EinsteinFrame}
\end{eqnarray}
where 
\begin{eqnarray}
V(\phi)&=&\frac{1}{2\kappa^2}\left[\frac{A}{F'(A)} - 
\frac{F(A)}{F'(A)^2}\right]\nonumber\\
&=&\frac{1}{2\kappa^2}\left\{\mathrm{e}^{\left(\sqrt{2\kappa^2/3}\right)\phi}
R\left(\mathrm{e}^{-\left(\sqrt{2\kappa^2/3}\right)\phi}\right)
-\mathrm{e}^{2\left(\sqrt{2\kappa^2/3}\right) 
\phi}F\left[R\left(\mathrm{e}^{-\left(\sqrt{2\kappa^2/3}\right)\phi}\right)\right]
\right\}\,.\label{V}
\end{eqnarray}
Here, $R(\mathrm{e}^{-\sqrt{2\kappa^2/3}\phi})$ is the solution of
(\ref{sigma}) with $A=R$ in terms of the field,
the Ricci scalar $\tilde{R}$ is evaluated respect to the conformal metric
$\tilde{g}_{\mu\nu}$, and $\tilde{g}= \mathrm{e}^{-4\sqrt{2\kappa^2/3}\phi}g$ is the determinant of such a metric\footnote{
In this paper, we will omit the tilde to denote all the quantities evaluated 
in the Einstein frame.}.\\
\\
Let us analyze inflation in the two different frameworks here presented. To do it, we will consider an ``unified'' formalism from which we derive the single cases. We start from the following Lagrangian,
\begin{equation}
\mathcal L= \frac{F(R)}{2\kappa^2}-\frac{g^{\mu\nu}\partial_{\mu}\phi\partial_{\nu}\phi}{2}-V(\phi)\,,\label{L}
\end{equation}
from which one easily gets (\ref{action}) and (\ref{EinsteinFrame}).

In a flat Friedmann-Robertson-Walker (FRW) universe with metric
\begin{equation}
ds^2=-dt^2+a(t)^2 {\bf dx}^2\,,\label{metric}
\end{equation}
where $a\equiv a(t)$ is the scale factor,
the Equations of Motion (EOMs) are obtained as
\begin{equation}
\frac{3 F_R (R) H^2}{\kappa^2}=\frac{\dot\phi^2}{2}+V(\phi)+\frac{1}{2\kappa^2}\left(R F_R(R)-F(R)\right)-\frac{3 H\dot F_R (R)}{\kappa^2}\,,\label{EOM1u}
\end{equation}
\begin{equation}
-\frac{2 F_R(R)\dot H}{\kappa^2}=\dot\phi^2+\frac{\ddot F_R (R)}{\kappa^2}-\frac{H \dot F_R (R)}{\kappa^2}\,.\label{EOM2u}
\end{equation}
Here, $H=\dot a/a$ is the Hubble parameter, the dot being the time derivative, and
\begin{equation}
F_R(R)=\frac{d F(R)}{d R}\,.\label{Fprime}
\end{equation}
Moreover, the continuity equation for the scalar field is given by
\begin{equation}
\ddot\phi+3H\dot\phi=-\frac{d V(\phi)}{d\phi}\,.\label{consu}
\end{equation}
The Hubble parameter and therefore the curvature are almost constant during inflation and one expects that the magnitude of the ``slow-roll'' parameters~\cite{sr},
\begin{equation}
\epsilon_1=-\frac{\dot H}{H^2}\,,\quad\epsilon_2=\frac{\ddot\phi}{H\dot\phi}\,,\quad
\epsilon_3=\frac{\dot F_R(R)}{2 H F_R(R)}\,,\quad
\epsilon_4=\frac{\dot E}{2 H E}\,,\label{srpar}
\end{equation}
where
\begin{equation}
E=\frac{F_R(R)}{\kappa^2}+\frac{3\dot F_R(R)^2}{2\dot\phi^2\kappa^2}\,,
\end{equation}
is very small (``slow-roll approximation''). In particular, it must be $0<\epsilon_1\ll 1$ to have acceleration. Under the slow-roll approximation, Eq.~(\ref{EOM1u}) and Eq.~(\ref{consu}) read
\begin{equation}
\frac{3 F_R(R) H^2}{\kappa^2}\simeq V(\phi)+\frac{1}{2\kappa^2}\left(R F_R(R)-F(R)\right)\,,
\quad
3H\dot\phi\simeq-\frac{d V(\phi)}{d\phi}\,.\label{srEOMs}
\end{equation}
In pure scalar field theory with $F(R)=R$, we recover
\begin{equation}
\frac{3 H^2}{\kappa^2}\simeq V(\phi)\,,
\quad
3H\dot\phi\simeq-\frac{d V(\phi)}{d\phi}\,.\label{srEOMs2}
\end{equation}
The spectral index and the tensor-to-scalar ratio for the general model (\ref{L}) are given by~\cite{corea},
\begin{equation}
n_s=1-4\epsilon_1-2\epsilon_2+2\epsilon_3-2\epsilon_4\,,\quad
r=16(\epsilon_1+\epsilon_3)\,,
\end{equation}
where $|\epsilon_{1,2,3,4}|\ll 1$ in the slow-roll regime. This indexes measure the perturbations at the end of inflation and must be confronted with the Planck data in (\ref{viablespectral}).\\
\\
In pure scalar field theory with $F(R)=R$, by using (\ref{srEOMs2}), one has 
\begin{equation}
\epsilon_1\simeq\frac{1}{2\kappa^2 V(\phi)^2}\left(\frac{d V(\phi)}{d\phi}\right)^2\,,\quad\epsilon_2\simeq\frac{\dot\epsilon_1}{H\epsilon_1}\simeq\epsilon_1-\frac{1}{\kappa^2 V(\phi)}\left(\frac{d^2 V(\phi)}{d\phi^2}\right)\,,
\quad \epsilon_2=\epsilon_4=0\,.\label{slowparscalar}
\end{equation}
As a consequence, we obtain
\begin{equation}
n_s\simeq 1-6\epsilon+2\eta\,,\label{spectralscalar}
\end{equation}
with 
\begin{equation}
\epsilon\equiv\epsilon_1=\frac{1}{2\kappa^2 V(\phi)^2}\left(\frac{d V(\phi)}{d\phi}\right)^2\,,\quad \eta\equiv\epsilon_1-\epsilon_2=\frac{1}{\kappa^2 V(\phi)}\left(\frac{d^2 V(\phi)}{d\phi^2}\right)\,.\label{spectralscalar2}
\end{equation}
In this case, the tensor-to-scalar ratio simply leads to
\begin{equation}
r\simeq 16\epsilon\,.\label{tensorscalar}
\end{equation}
On the other hand, in pure modified gravity with $\phi=0$ and $V(\phi)=0$, given that $\epsilon_2=0$, $\epsilon_1\simeq -\epsilon_3(1-\epsilon_4)$, since in the slow-roll approximation $\epsilon_1\simeq-\epsilon_3$ and $\epsilon_4\simeq-3\epsilon_1+\dot\epsilon_1/(H\epsilon_1)$, we obtain
\begin{equation}
n_s\simeq 1-6\epsilon_1-2\epsilon_4=1-\frac{2\dot\epsilon_1}{H\epsilon_1}\,,\quad
r\simeq 48\epsilon_1^2\,,\label{spectralMG}
\end{equation} 
where for the tensor-to-scalar ratio we used the second order corrections, since at the first order $r=0$.
In the next section, we will review a class of viable inflation in $F(R)$-modified gravity in its two different frameworks by starting from the Planck results (\ref{viablespectral}).

\section{Viable Starobinsky-like inflation}

In both scalar field and modified gravity theories, the equations of motion can be written as
\begin{equation}
\frac{3H^2}{\kappa^2}=\rho_\text{eff}\,,\quad 
-\frac{(3H^2+2\dot H)}{\kappa^2}=p_\text{eff}\,,\label{EOMsgen}
\end{equation}
where $\rho_\text{eff}\,,p_\text{eff}$ are the effective energy density and pressure of the universe (in the case of modified gravity they contain gravitational terms) satisfying a continuity equation
\begin{equation}
\dot\rho_\text{eff}+3H(\rho_\text{eff}+p_\text{eff})=0\,.\label{due}
\end{equation}
In the specific, for scalar theories one has
\begin{equation}
\rho_\text{eff}=\frac{\dot\phi^2}{2}+V(\phi)\,,\quad p_\text{eff}=\frac{\dot\phi^2}{2}-V(\phi)\,,
\label{rhopscalar}
\end{equation}
while for $F(R)$-modified gravity we find
\begin{eqnarray}
\rho_{\mathrm{eff}} &=& \frac{1}{2\kappa^2}\biggl[(RF_{R}(R)
-F(R))-6H\dot{F}_{R}(R)
-6H^{2}\left(F_R(R)-1\right)
\biggr]\,,\nonumber\\
p_{\mathrm{eff}} &=& \frac{1}{2\kappa^2}\biggl[
(F(R)-R F_{R}(R))+4H\dot{F}_{R}(R)+2\ddot{F}_{R}(R)
+(4\dot{H}+6H^{2})\left(F_R(R)-1\right)                      
\biggr]\,.\label{rhopMG}
\end{eqnarray}
An effective Equation of State (EoS) can be introduced as
\begin{equation}
p_\text{eff}=\omega_\text{eff}\rho_\text{eff}\,,\label{tre}
\end{equation}
with $\omega_\text{eff}$ an effective EoS parameter.
At the beginning of inflation $\omega_\text{eff}$ must be close to minus one
but not vanishing in order to exit from early-time acceleration~\cite{muk1, muk2}. Moreover, one may require $-1<\omega_\text{eff}$ to avoid $\omega_\text{eff}=-1$ at some time of inflation, since the pure de Sitter solution can be a final attractor of the system. Acceleration vanishes when $-1/3\leq \omega_\text{eff}$, namely
the ``strong energy condition'' is violated. Thus,
a reasonable Ansatz for the effective EoS parameter of inflation in terms of the $e$-folds (\ref{efolds}) may be
~\cite{muk1},
\begin{equation}
1+\omega_\text{eff}\simeq\frac{\beta}{(N+1)^\alpha}\,,\quad 0<\alpha\,,\beta\,.\label{Anomega}
\end{equation}
Here, $\beta$ is a number on the order of the unit. As a consequence, from (\ref{due}) and (\ref{tre}), by taking into account that $d/dt=-H(t) d/d N$, one derives
\begin{equation}
\rho_\text{eff}\simeq\rho_\text{f}(N+1)^{3\beta}\,,\quad\alpha=1\,,\label{rho1}
\end{equation}
\begin{equation}
\rho_\text{eff}\simeq\rho_0\exp\left[-\frac{3\beta}{(\alpha-1)(N+1)^{\alpha-1}}\right]\,,\label{rho2}
\quad 1\neq \alpha\,,
\end{equation}
where $\rho_{0,\text{f}}$ are integration constants: in particular, $\rho_\text{f}$ is the effective energy density at the end of inflation at $N=0$ in the case of $\alpha= 1$, and $\rho_0$ is the effective energy density at the beginning of inflation at $1\ll N$ in the case of $1<\alpha$. By starting from this results, we can now reconstruct the spectral index and the tensor-to-scalar ratio which realize (\ref{Anomega}) and the corresponding models in the different representations.\\
\\
In scalar field theories, by using (\ref{slowparscalar}) with (\ref{EOMsgen})--(\ref{rhopscalar}) and (\ref{tre})--(\ref{Anomega}) one has in slow-roll approximation (\ref{srEOMs2}),
\begin{equation}
\epsilon_1=\frac{3(1+\omega_\text{eff})}{2}\simeq\frac{3\beta}{2(N+1)^\alpha}\,,\quad
\epsilon_2=-\frac{d\ln [1+\omega_\text{eff}]}{d N}\simeq\frac{\alpha}{N+1}\,.
\end{equation}
Thus, the spectral index and the tensor-to-scalar 
ratio in (\ref{spectralscalar})--(\ref{tensorscalar}) result to be
\begin{equation}
n_s\simeq 1-\left[\frac{3\beta+\alpha(\mathcal N+1)^{\alpha-1}}{(\mathcal N+1)^{\alpha}}\right]\,,
\quad r\simeq \frac{24\beta}{(\mathcal N+1)^{\alpha}}\,.
\end{equation}
where the slow-roll parameters have beed evaluated during inflation at $N=\mathcal N$, with $\mathcal N$ the total $e$-folds number defined in (\ref{Nfolds}).
 
Therefore, by looking for the last Planck data in (\ref{viablespectral}), only the case $\alpha=2$ with $\mathcal N\simeq 60$
leads to viable values for the spectral index and the tensor-to-scalar ratio. The case $\alpha=1$ is also quite interesting, since it corresponds to power-law scalar potential~\cite{muk1},
but, despite to the fact that it gives a correct value of the spectral index, the tensor-to-scalar ratio is in general larger than the Planck result.

For $\alpha=2$ the EoS parameter (\ref{Anomega}) with (\ref{rho2}) reads
\begin{equation}
\omega_\text{eff}\simeq -1+\frac{1}{9\beta}\log\left[\frac{\rho_\text{eff}}{\rho_0}\right]^2\,,\label{omegaalpha2}
\end{equation}
and from (\ref{tre}) with (\ref{rhopscalar})  in the slow-roll approximation $\dot\phi^2\ll V(\phi)$
it is easy to reconstruct
\begin{equation}
\dot\phi\simeq\frac{\sqrt{V(\phi)}}{3\sqrt{\beta}}\log\left[\frac{\rho_0}{V(\phi)}\right]\simeq
\frac{\sqrt{V(\phi)}}{3\sqrt{\beta}}\left(\frac{\rho_0}{V(\phi)}-1\right)\,,
\end{equation}
where we considered $V(\phi)$ close to $\rho_0$ during inflation.
Thus, from the second expression in (\ref{srEOMs2}) we get
\begin{equation}
V(\phi)=\rho_0\left(1-c_1 \text{e}^{\sqrt{\kappa^2/(3\beta)}\phi}
\right)\,,\quad 0<c_1\,,\label{VStarStar}
\end{equation}
where $c_1$ is a constant whose positivity comes from the implicit assumption $0<\dot\phi$. Inflation takes place in the limit $\phi\rightarrow-\infty$, which brings to the slow-roll approximation in (\ref{srEOMs2}). This is the form of the potential in chaotic inflation which leads to a viable accelerated expansion with spectral index and tensor-to-scalar ratio
\begin{equation}
n_s\simeq 1-\frac{2}{\mathcal N+1}\,,\quad r\simeq \frac{24\beta}{\left(\mathcal N+1\right)^2}\,,
\label{spectralstaro}
\end{equation}
where $55<\mathcal N<65$. We will return later on this result.\\
\\
In $F(R)$-modified gravity one has from (\ref{spectralMG}) with the Ansatz (\ref{Anomega}),\begin{equation}
n_s\simeq 1-\frac{2\alpha}{(\mathcal N+1)}\,,
\quad r\simeq\frac{108\beta^2}{(\mathcal N+1)^{2\alpha}}\,,
\end{equation}
such that the choice $\alpha=1$ with $\mathcal N\simeq 60$ satisfies the Planck results in (\ref{viablespectral}). The corresponding $\omega_\text{eff}$ parameter is given by
\begin{equation}
\omega_\text{eff}=-1+\beta\left(\frac{\rho_\text{f}}{\rho_\text{eff}}\right)^{\frac{1}{3\beta}}\,,
\end{equation}
where we have used (\ref{rho1}). For the sake of simplicity, in the following we will set $\beta=1/3$. Thus, by plugging (\ref{rhopMG}) in (\ref{tre}) we obtain
\begin{equation}
4\dot H \left(F_R(R)-1\right)-2H\dot F_R(R)+2\ddot F_R(R)=\frac{2\kappa^2\rho_\text{f}}{3}\,.
\label{omcondm}
\end{equation}
One also has
\begin{equation}
R=12H^2+6\dot H\,,\quad
H=\sqrt{\frac{\kappa^2\rho_\text{f}}{3}}\sqrt{N+1}\,,\quad\dot H=-\frac{\kappa^2\rho_\text{f}}{6}\,,\label{solpart}
\end{equation}
where we have expressed the Hubble parameter and its time derivative in terms of the $e$-fold left to the end of inflation. In this way, from (\ref{omcondm}) we get
\begin{equation}
-F_R(R)+\left(\frac{2N+3}{2}\right)\frac{d F_R(R)}{d N}+(N+1)\frac{d^2 F_R(R)}{d N^2}=0\,,\quad F_R(R)=c_0\left(\frac{3}{2}+N\right)\,,
\end{equation}
$c_0$ being an integration constant.
By making use of the relation $R=\kappa^2\rho_\text{f}(4N+3)$, one has
\begin{equation}
F_R(R)=\frac{c_0 R}{2\rho_\text{f}}+\frac{3c_0\kappa^2}{2}
\,,\quad
F(R)=\frac{c_0 R^2}{4\rho_\text{f}}+\frac{3c_0\kappa^2 R}{2}
+\lambda\,,\label{eqexf}
\end{equation}
with $\lambda$ a ``cosmological constant'' fixed by the first EOM in (\ref{EOMsgen}) with (\ref{rhopMG}) as
\begin{equation}
\lambda=\frac{c_0\kappa^4\rho_\text{f}}{4}\,.\label{cc}
\end{equation}
If we want to recover the Hilbert-Einstein term of General Relativity we must set
$c_0=2/(3\kappa^2)$ and the model
\begin{equation}
F(R)=R+\frac{R^2}{6\kappa^2\rho_\text{f}}+\frac{\rho_\text{f}\kappa^2}{6}\,,\label{mnew}
\end{equation}
realizes viable inflation in agreement with the Planck satellite data. We note that in deriving this result we did not use the slow-roll approximation in the EOMs as in (\ref{srEOMs}).\\
\\
Some remarks are in order about the potential in (\ref{VStarStar}) and the gravitational model in (\ref{mnew}). During inflation\footnote{This condition implies that the term $R^2$ is much bigger than the Hilbert-Einstein contribute to the action, such that $F(R)/(2\kappa^2)\sim R^2$. The $R^2$-model possesses the de Sitter solution for an arbitrary boundary value of the curvature, like in the chaotic scalar field inflation. We also mention that during inflation the mass encoded in $\kappa^2=8\pi/M^2$ is not expected to have the value of the Planck Mass (see last section), and curvature remains in sub-planckian domain.}
\begin{equation}
\kappa^2\rho_\text{f}\ll R\,,
\end{equation}
such that the $F(R)$-model (\ref{mnew}) can be rewritten as
\begin{equation}
F(R)\simeq R+\frac{R^2}{6\kappa^2\rho_\text{f}}\,.\label{Starob}
\end{equation}
The scalar field representation of such a model is given by (\ref{sigma})--(\ref{EinsteinFrame}) with the potential 
\begin{equation}
V(\phi)=\frac{3\rho_\text{f}}{4}\left(1-\text{e}^{\sqrt{2\kappa^2/3}\phi}
\right)^2\,,\label{VStarob}
\end{equation}
which corresponds to (\ref{VStarStar}) with\footnote{The scale of the inflation in the Einstein frame (EF) is different from the one in the Jordan frame (JF). In particular, for the constant curvature of de Sitter space-time, 
$R_{\text{JF}}= \text{e}^{-\sqrt{2\kappa^2/3}\phi}R_{\text{EF}}$, 
such that $R_{\text{EF}}\ll R_{\text{JF}}$ when $\phi\rightarrow-\infty$. This is the reason for which $\rho_\text{f}$, namely the effective energy density at the end of inflation, goes into $\rho_0$, namely the energy density at the beginning of inflation, when we pass from Jordan- to Einstein-frame.}
 $\rho_0=3\rho_\text{f}/4$, $c_1=2$, $\beta=1/2$ in the limit $\phi\rightarrow-\infty$. 
Inflation in cosmological models with Lagrangian $R+c_1 R^2$ and the demonstration of the conformal equivalence of such theories and General Relativity plus a scalar field with the asymmetric potential above was first derived in Ref.~\cite{superBarrow}.
The model (\ref{Starob}) with its scalar field representation (\ref{VStarob})
is usually called ``Starobinsky model'' and leads to the spectral index and the tensor-to-scalar ratio (\ref{spectralstaro}) with $\beta=1/2$, namely\footnote{We should remember that the $\beta$-parameter in the reconstruction of scalar field models is different from the $\beta$-parameter in the reconstruction of $F(R)$-models, the last one being fixed at $\beta=1/3$.}
\begin{equation}
n_s=1-\frac{2}{\mathcal N+1}\,,\quad r=\frac{12}{\left(\mathcal N+1\right)^2}\,.
\label{spectralstaro2}
\end{equation}
Some extensions by including a cosmological constant in the Jordan frame or, equivalently, a (negligible) term proportional to $\sim\exp\left[2\sqrt{\kappa^2/(3\beta)}\phi\right]$ in the potential of the Einstein frame are possible, while with the setting (\ref{cc}) we obtain an exact
accelerated solution. Viable $F(R)$-inflation under the Ansatz (\ref{Anomega}) may be found also for $\beta\neq 1/3$, but, since $\beta$ is a parameter on the order of the unit, we still expect to remain near to the Starobinsky inflation in the two frameworks. In the next section, we will investigate this cases by using a different approach.

\section{Reconstruction of $F(R)$-inflation from scalar field representation}

In this section we would like to investigate
several generalizations of the potential (\ref{VStarStar}) to reproduce inflation according with the Planck data. $F(R)$-gravity in the corresponding Jordan framework will be reconstruct
(for reconstruction of $F(R)$-inflation in the original $F(R)$-frame, see Ref.~\cite{RecOd}).
In order to do it, we divide 
Eq.~(\ref{V}) to 
$\exp\left[2\sqrt{2\kappa^2/3}\right]$, and we take the derivative with respect to 
$R$,
\begin{equation}
R 
F_R(R)=-2\kappa^2\sqrt{\frac{3}{2\kappa^2}}\frac{d}{d\phi}\left[\frac{V(\phi)}{e^{2\left(\sqrt{2\kappa^2/3}\right)\phi}}\right]\,.\label{start}
\end{equation}
Giving the explicit form of the potential $V(\phi)$ with relation (\ref{sigma}), we can solve such an equation respect to $R$ and therefore recover
the $F(R)$-gravity model in the Jordan framework.  In this process, one acquires 
the integration constant, which can be fixed by  using Eq.~(\ref{V}).

\subsubsection*{$V(\phi)\sim \left[c_0-c_1\exp[\kappa\phi]+c_2\exp[2\kappa\phi]\right]/(2\kappa^2)$: $R+ 
R^2/(4c_0)+\lambda $-models\label{4.1}}

Let us start by the minimal generalization of the Starobinsky potential
\begin{equation}
V(\phi)=\frac{1}{2\kappa^2}\left[c_0-c_1\text{e}^{\sqrt{2\kappa^2/3}\phi}+c_2\text{e}^{2\sqrt{2\kappa^2/3}\phi}
\right]\,,\quad 0<c_0\,,c_1\,,\label{cucu}
\end{equation}
with $c_{0,1,2}$ dimensional constants ($[c_{0,1,2}]=[\kappa^{-2}]$).
Equation (\ref{start}) leads to
\begin{equation}
2c_0F_R^2-c_1F_R-RF_R=0\,,\quad
F_R(R)=\frac{c_1}{2c_0}+\frac{R}{2c_0}\,,\label{FF}
\end{equation}
such that one derives the model
\begin{equation}
F(R)=\frac{c_1}{2c_0}R+\frac{R^2}{4c_0}+\lambda\,,
\quad
\lambda=\frac{c_1^2}{4c_0}-c_2\,,\label{Staro2}
\end{equation}
where the cosmological constant $\lambda$ has been fixed by Eq.~(\ref{V}).
In order to have the Hilbert-Einstein 
term of General Relativity, we must put $c_1/(2c_0)=1$, namely
\begin{equation}
F(R)=R+\frac{R^2}{4c_0}+c_0-c_2\,.\label{Staro3}
\end{equation}
For $c_0=c_2$ we recover the Starobinsky model (\ref{Starob}) with the corresponding potential (\ref{VStarob}) after the identification $c_0=1/(3\rho_\text{f})$.
We also mention that the choice $c_0=0$ brings the model to exhibit a particular class of static 
spherically symmetric solutions~\cite{uno}.

In the Einstein frame, inflation starts at large and negative values of the field, when the EOMs with the slow-roll approximation (\ref{srEOMs2}) lead to
\begin{equation}
H^2\simeq\frac{\kappa^2 V(\phi)}{3}\simeq\frac{\gamma}{12}\,,
\quad
3H\dot\phi\simeq -\frac{d V(\phi)}{d \phi}\simeq
\left(\frac{\gamma}{\sqrt{6\kappa^2}}\right)\text{e}^{\sqrt{2\kappa^2/3}\phi}\,,
\quad\gamma=2 c_0\label{copy}\,.
\end{equation}
Thus, a (quasi) de Sitter solution with $H^2\propto\gamma$ is realized and the field slowly moves to the minimum of the potential at 
$V(0)=-\gamma/(4\kappa^2)+c_2/(2\kappa^2)$ when inflation ends.
During inflation the slow-roll parameters (\ref{spectralscalar2}) read
\begin{equation}
\epsilon\simeq\frac{4}{3}\frac{1}{(2-\text{e}^{-\sqrt{2\kappa^2/3}\phi})^2}\,,\quad\eta\simeq\frac{4}{3}\frac{1}{\left(2-\text{e}^{-\sqrt{2\kappa^2/3}\phi}\right)}\,,\label{star}
\end{equation}
and are small in the limit $\phi\rightarrow-\infty$. The $e$-folds number (\ref{Nfolds}) which measures the total amount of inflation reads
\begin{equation}
\mathcal N\equiv\ln \left[\frac{a(t_\text{f})}{a(t_\text{i})}\right]=\int_{t_\text{i}}^{t_\text{f}} H dt\simeq 
\kappa^2\int^{\phi_0}_{\phi_\text{f}}\frac{V(\phi)}{dV(\phi)/d\phi}d\phi\,,
\end{equation}
with $\phi_{0,\text{f}}$ the values of the field at the beginning and at
the end of inflation, as usually. In our case we obtain 
\begin{equation}
\mathcal N\simeq\frac{3\text{e}^{-\sqrt{2\kappa^2/3}\phi}}{4}\Big\vert^{\phi_0}_{\phi_\text{f}}\simeq\frac{3\text{e}^{-\sqrt{2\kappa^2/3}\phi_0}}{4}\,,
\end{equation}
where we have taken into account the fact that the field is negative and $|\phi_\text{f}|\ll |\phi_0|$. As a consequence, during inflation 
\begin{equation}
\epsilon\simeq\frac{3}{4\mathcal N^2}\,,\quad\eta\simeq-\frac{1}{\mathcal N}\,,
\end{equation}
and one derives for the spectral index and the tensor-to-scalar ratio in (\ref{spectralscalar}) and (\ref{tensorscalar}),
\begin{equation}
n_s\simeq 1-\frac{2}{\mathcal N}\,,\quad r\simeq\frac{12}{\mathcal N^2}\,, 
\end{equation}
namely (\ref{spectralstaro2}) with $1\ll \mathcal N$: we have already seen that this expressions are in agreement with the last Planck satellite data (\ref{viablespectral}). 
Thus, the behaviour of the scalar field model with potential (\ref{cucu}) brings to the same results of Starobinsky inflation, eventually with a
minimum of the potential different to zero and a cosmological constant in the Jordan frame\footnote{The appearance of a cosmological constant in the Jordan frame depends on the term  
$\propto\exp[2\sqrt{2\kappa^2/3}\phi]$ in the potential of the Einstein frame, and does not 
modify the behaviour of the models during the early-time 
acceleration: however, it leads to a different minimum of the potential where the field falls at the end of inflation.}.

\subsubsection*{$V(\phi)\sim\left[3\gamma/4-\gamma\exp[\kappa\phi/2]\right]/\kappa^2$: $R/2+c_1 R^2+c_2 
(R+R_0)^{3/2}$-models}

In this subsection we would like to propose a potential with a different behaviour respect to the 
Starobinsky one (which 
decreases as $\sim\exp[-\kappa\phi]$). We consider the following form of $V(\phi)$,
\begin{equation}
V(\phi)=\frac{\alpha}{\kappa^2}-\frac{\gamma}{\kappa^2}\text{e}^{\sqrt{2\kappa^2/3}\phi/2}\,,\quad 0<\alpha\,,\gamma\,\label{V2}
\end{equation}
with $\alpha,\gamma$ dimensional constants ($[\alpha,\beta]=[\kappa^{-2}]$). It follows
from (\ref{start}) the (real) solution
\begin{equation}
F_R(R)=\frac{9\gamma^2+8R\alpha+3\sqrt{16 
R\alpha\gamma^2+9\gamma^4}}{32\alpha^2}\,.
\end{equation}
Since at small curvature ($R\ll \alpha\,,\gamma$) we would like
to find the Einstein's gravity ($F_R(R)=1$), we have to set
$\alpha=3\gamma/4$,
\begin{equation}
F_R(R)=\frac{1}{2}+\frac{1}{3\gamma}R+\frac{\sqrt{3}}{6}\sqrt{4R/\gamma+3}\,,
\quad
F(R)=\frac{R}{2}+\frac{R^2}{6\gamma}+\frac{\sqrt{3}}{36}\left(4R/\gamma+3\right)^{3/2}+\lambda\,,\quad \lambda=\frac{\gamma}{4}\,,\label{zippozap}
\end{equation}
where $\lambda$ has been fixed by using (\ref{V}). In this way,
$F(R\ll\gamma)\simeq R+\gamma/2$. We also may cancel the cosmological constant if we put $\lambda=-\gamma/4$ and we add the term $-(\gamma/(4\kappa^2))\exp\left[2\sqrt{2\kappa^2/3}\phi\right]$ inside the potential (\ref{V2}): such a term will not modify the behaviour of the model during inflation and will not be considered in the following.
Thus, potential 
(\ref{V2}) finally reads
\begin{equation}
V(\phi)=\frac{\gamma}{\kappa^2}\left(\frac{3}{4}-\text{e}^{\sqrt{2\kappa^2/3}\phi/2}
\right)\,.
\end{equation}
The value of the field during inflation is negative and very large. From (\ref{srEOMs2}) with $\phi\rightarrow-\infty$ we derive
\begin{equation}
H^2\simeq\frac{\gamma}{4}\,,
\quad
3H\dot\phi\simeq 
\left(\frac{\gamma}{\sqrt{6\kappa^2}}\right)\text{e}^{\sqrt{2\kappa^2/3}\phi/2}\,,
\end{equation}
and the slow roll-parameters (\ref{spectralscalar2}) read
\begin{equation}
\epsilon\simeq\frac{4}{3}\frac{1}{\left(4-3\text{e}^{-\sqrt{2\kappa^2/3}\phi/2}\right)^2}\,,\quad\eta\simeq\frac{2}{3}\frac{1}{\left(4-3\text{e}^{-\sqrt{2\kappa^2/3}\phi/2}\right)}\,.
\end{equation}
The total e-folds is given by
\begin{equation}
\mathcal N\simeq\kappa^2\int^{\phi_0}_{\phi_\text{f}}\frac{V(\phi)}{dV(\phi)/d\phi}d\phi\simeq\frac{9\text{e}^{-\sqrt{2\kappa^2/3}\phi_0/2}}{2}\,,
\end{equation}
such that during inflation
\begin{equation}
\epsilon\simeq\frac{3}{\mathcal N^2}\,,\quad 
\eta\simeq-\frac{1}{\mathcal N}\,,
\end{equation}
and from (\ref{spectralscalar})--(\ref{tensorscalar}) we get
\begin{equation}
n_s\simeq1-\frac{2}{\mathcal N}\,,\quad r\simeq\frac{48}{\mathcal N^2}\,.
\end{equation}
These data are compatible with the Planck results (\ref{rangeN})--(\ref{viablespectral}) and lead to a viable expansion of the early-time universe.

\subsubsection*{$\left[V(\phi)\sim\gamma(2-n)/2-\gamma\exp[n\kappa\phi]\right]/\kappa^2\,,0<n<2$: $c_1 
R^2+c_2 R^{2-n}$-models}

Here we investigate the feature of the general
potential
\begin{equation}
V(\phi)=\frac{\alpha}{\kappa^2}-\frac{\gamma}{\kappa^2}\text{e}^{n\sqrt{2\kappa^2/3}\phi}\,,
\quad 0<\alpha\,,\gamma\,,
\quad 0<n<2\,,\label{V3}
\end{equation}
where $\alpha\,,\gamma$ are dimensional constants again ($[\gamma\,,\alpha]=[\kappa^{-2}]$) and $n$ is a positive number smaller than two. From equation (\ref{start}) we derive
\begin{equation}
F_R(R)+\frac{\gamma}{2\alpha}F_R(R)^{1-n}(n-2)=\frac{R}{4\alpha}\,.\label{pippo2}
\end{equation}
At the perturbative level, it is possible to see~\cite{uno} that, by choosing
\begin{equation}
\alpha=\frac{\gamma(2-n)}{2}\,,\label{alphaalpha}
\end{equation}
at small curvature the solution of (\ref{pippo2}) reads
\begin{equation}
F_R(R\ll\gamma)\simeq 1+c_1 R+c_2 R^2+c_3 R^3+...\,,\label{espansione}
\end{equation}
where $c_{1,2,3,...}$ are given by a recursive formula,
\begin{equation}
c_1=\frac{1}{2\gamma\,n(2-n)}\,,\quad
c_2=-\frac{n c_1^2}{2}\,,\quad
c_3=\frac{(1-n)(n+1)c_1^3}{6}\,, ...
\end{equation}
when $|c_1 R|\,,|c_2 R^2|\,, |c_3 R^3|...\ll 1$. Since
$c_1\propto\gamma^{-1}\,,c_2\propto\gamma^{-2}\,,c_3\propto\gamma^{-3}...$, in the considered limit $R\ll \gamma$ the model becomes a (negligible) higher order curvature correction of General Relativity. In the case $n=1/2$ we recover the expansion of (\ref{zippozap}) when $R\ll\gamma$. We mention that a cosmological constant may emerge in the Jordan frame, but it is always possible to cancel it by introducing a term proportional to $\sim\exp\left[2\sqrt{2\kappa^2/3}\phi\right]$ in the corresponding potential of the Einstein-frame.

When $\gamma\ll R$, the asymptotic solution of 
Eq.~(\ref{pippo2}) with (\ref{alphaalpha}) is derived as\footnote{
See footnote [3]. In the Einstein frame inflation is realized at $R_{\text{EF}}\sim\gamma$, but in the Jordan frame $\gamma\ll R_\text{JF}$.}
\begin{eqnarray}
F_R(R\gg\gamma)&\simeq&\left(\frac{R}{2\gamma(2-n)}\right)+\left(\frac{R}{2\gamma(2-n)}\right)^{1-n}\,,\nonumber\\\nonumber\\
F(R\gg\gamma)&\simeq&\frac{1}{2}\left(\frac{R^2}{2\gamma(2-n)}\right)+\frac{1}{2-n}\left(\frac{1}{2\gamma(2-n)}\right)^{1-n}R^{2-n}\,.
\label{megastana}
\end{eqnarray}
For $n=1/2$ we find the expansion of (\ref{zippozap})  when $\gamma\ll R$.
As another example, we may consider the case $n=1/3$. The reconstruction leads 
to the following exact form of the model in the Jordan frame,
\begin{equation}
F_R(R)=\frac{1}{3}+\frac{3}{10}\left(\frac{R}{\gamma}\right)
+\frac{2}{3\times 
5^{1/3}}\left[9\left(\frac{R}{\gamma}\right)+5\right]\frac{1}{\Delta^{1/3}}+\frac{1}{6\times
5^{2/3}}\Delta^{1/3}\,,
\end{equation}
with
\begin{equation}
\Delta=200+243\left(\frac{R}{\gamma}\right)^2+540\left(\frac{R}{\gamma}\right)+27\sqrt{81\left(\frac{R}{\gamma}\right)^4+40\left(\frac{R}{\gamma}\right)^3}\,.
\end{equation}
In the limit $R\ll\gamma$ we can write
\begin{equation}
F_R(R\ll\gamma)\simeq1+\frac{9}{10}\left(\frac{R}{\gamma}\right)+...\,,
\end{equation}
which corresponds to (\ref{espansione}) with $n=1/3$.
Therefore, in the limit $\gamma \ll R$, one has
\begin{equation}
F_R(R\gg\gamma)\simeq 
\frac{3}{10}\left(\frac{R}{\gamma}\right)+\left(\frac{3}{10}\right)^{\frac{2}{3}}\left(\frac{R}{\gamma}\right)^{\frac{2}{3}}\,,
\end{equation}
which corresponds to (\ref{megastana}) with $n=1/3$.
We also note that the model (\ref{FF}) is analogous to the case $n=1$.

Let us analyze chaotic inflation in the Einstein frame. By plugging (\ref{alphaalpha})  in (\ref{V3}) we obtain
\begin{equation}
V(\phi)=\frac{\gamma(2-n)}{2\kappa^2}-\frac{\gamma}{\kappa^2}\text{e}^{n\sqrt{2\kappa^2/3}\phi}\,.\label{n<2}
\end{equation}
The EOMs in (\ref{srEOMs2}) in the limit $\phi\rightarrow-\infty$  read
\begin{equation}
H^2\simeq\frac{\gamma(2-n)}{6}\,,
\quad
3H\dot\phi\simeq 
\left(\frac{n\gamma\sqrt{2}}{\sqrt{3\kappa^2}}\right)\text{e}^{n\sqrt{2\kappa^2/3}\phi}\,,
\end{equation}
and inflation ends when the field reaches the minimum of the potential at $V(\phi=0^-)= -n\gamma/(2\kappa^2)$.
The slow-roll parameters (\ref{spectralscalar2}) are given by
\begin{equation}
\epsilon\simeq\frac{4n^2}{3}\frac{1}{\left(2+(n-2)\text{e}^{-n\sqrt{2\kappa^2/3}\phi}\right)^2}\,,
\quad
\eta\simeq\frac{4n^2}{3}\frac{1}{\left(2+(n-2)\text{e}^{-n\sqrt{2\kappa^2/3}\phi}\right)}\,.
\end{equation}
The total $e$-folds is derived as
\begin{equation}
\mathcal N\simeq\frac{3(2-n)\text{e}^{-n\sqrt{2\kappa^2/3}\phi_0}}{4n^2}\,,
\end{equation}
such that during inflation
\begin{equation}
\epsilon\simeq\frac{3}{4n^2\mathcal N^2}\,,\quad\eta\simeq-\frac{1}{\mathcal N}\,.
\end{equation}
As a consequence, the spectral index (\ref{spectralscalar}) and the tensor-to-scalar ratio (\ref{tensorscalar}) 
read
\begin{equation}
n_s\simeq1-\frac{2}{\mathcal N}\,,\quad r\simeq\frac{12}{n^2 \mathcal N^2}\,,
\end{equation}
and in order to satisfy the Planck data (\ref{viablespectral}) it must be
\begin{equation}
0.174<n<2\,.
\end{equation}
Thus, here and in the preceding subsection we have seen that the contribute of $R^{2-n}$-term with  $1/5<n<2$ in the gravitational action of the theory makes the chaotic $R^2$-inflation realistic and in agreement with the last Planck satellite data. The case $n=1$ corresponds to Starobinsky inflation, 
while different choices of $n$ in the given range modify the value of the tensor-to-scalar ratio during the early-time acceleration, but still lead to a viable anisotropy spectrum. In the slow curvature limit, the Lagrangian turns out to be the same of General Relativity. 

\section{Inflation from higher-derivative quantum gravity}

Modified gravity for inflation is mainly motivated by the fact that it may come from quantum effects at high curvature. In this section, we will briefly see how
the early-time acceleration in a 
general higher-derivative quantum gravity theory~\cite{B} can be realized. Here, the corrections to Einstein's gravity are encoded in the form of $F(R)$-gravity. 

By applying renormalization group (RG)
considerations one can find a RG improved effective gravitational action: this theory
is known to be multiplicatively-renormalizable theory and the
one-loop beta-functions which appear in the action are well-known and their asymptotically
free regime has been well investigated. In fact, by making use of
some techniques well-developed in quantum field theory for curved
spacetime~\cite{El} and by considering the sum of all leading logs of the theory, it is possible to derive such an effective action beyond the one-loop
approximation. 

The general form of a higher-derivative theory is
\begin{equation}
I=\int_\mathcal{M} d^4 x\sqrt{-g}\left(\frac{R}{\tilde\kappa^2}-\Lambda+a
R_{\mu\nu}R^{\mu\nu}+b R^2+c R_{\mu\nu\xi\sigma}R^{\mu\nu\xi\sigma}+d\Box
R\right)\,,\label{azione0}
\end{equation}
where $R_{\mu\nu}\,,R_{\mu\nu\xi\sigma}$ are 
the Ricci tensor and the Riemann tensor, respectively, and
$\Box\equiv g^{\mu\nu}\nabla_{\mu}\nabla_{\nu}$ is the
covariant d'Alembertian, ${\nabla}_{\mu}$ being the covariant derivative
operator associated with the metric.
If
$0<\tilde\kappa^2$ and $\Lambda\,,a\,,b\,,c\,,d$ are constants that characterize
the gravitational interaction, some terms can be drop down.
First of all, $\Box R$ is a surface
term and does not contribute to the dynamic of the system, and at second we can write
\begin{equation}
R_{\mu\nu}R^{\mu\nu}=\frac{C^2}{2}-\frac{G}{2}+\frac{R^2}{3}\,,\quad
R_{\mu\nu\xi\sigma}R^{\mu\nu\xi\sigma}=2C^2-G+\frac{R^2}{3}\,,
\end{equation}
where $G$ and $C^2$ are the Gauss-Bonnet term and the ``square'' of the Weyl
tensor,
\begin{equation}
G=R^2-4R_{\mu\nu}R^{\mu\nu}+R_{\mu\nu\xi\sigma}R^{\mu\nu\sigma\xi}\,,\quad
C^2=\frac{1}{3}R^2-2R_{\mu\nu}R^{\mu\nu}+R_{\xi\sigma\mu\nu}R^{\xi\sigma\mu\nu}\,.\label{Gauss}
\end{equation}
Thus, the contractions of the Riemann and the Ricci tensors can be replaced by the Gauss-Bonnet and the Weyl terms. Since the Gauss-Bonnet  is a topological invariant in four dimensions, and the Weyl tensor is identically zero on FRW metric (\ref{metric}), in inflationary scenario the theory reduces to $F(R)$-gravity.

In order to recover an asymptotically free theory by taking into account quantum effects, one has to use the
RG improved effective action,
 namely the coupling constants in (\ref{azione0}) are replaced 
by one-loop effective coupling
constants defined as 
log terms of
the characteristic mass scale of the theory (for detailed works see Refs.~\cite{B,El} and Ref.~\cite{rginfl}). 

The RG improved effective action read\footnote{When we introduce the effective running
constants in the higer derivative action (\ref{azione0}), we also get a contribution from the Gauss-Bonnet
and $\Box R$-terms. This fact will be discussed in the end of the section, but for the
moment we
    work with the simplified Lagrangian.}
\begin{equation}
I=\int_\mathcal{M}d^4\sqrt{-g}\left[\frac{R}{\kappa^2(t')}-\frac{\omega(t')}{3\lambda(t')}R^2+\frac{1}{\lambda(t')}C^2-\Lambda(t')\right]\,,
\label{actionQ}
\end{equation}
where the effective coupling constants $\lambda\equiv\lambda(t')$,
$\omega\equiv\omega(t')$, $\kappa^2\equiv\kappa^2(t')$ and
$\Lambda\equiv\Lambda(t')$ are solutions of the one-loop RG
equations~\cite{Fradkin},
\begin{eqnarray}
\frac{d\lambda}{d
t'}=-\beta_2\lambda^2\,,\quad
\frac{d\omega}{d
t'}=-\lambda(\omega\beta_2+\beta_3)\,,\quad
\frac{d\kappa^2}{d
t'}=\kappa^2\gamma\label{kappaeq}\,,\quad
\frac{d\Lambda}{d
t'}=\frac{\beta_4}{\left(\kappa^2\right)^2}-2\gamma\Lambda(t')
\label{couplingconst}\,,
\end{eqnarray}
with
\begin{eqnarray}
&&\beta_2=\frac{133}{10}\,,\quad\beta_3=\frac{10}{3}\omega^2+5\omega+\frac{5}{12}\,,\quad
\beta_4=\frac{\lambda^2}{2}\left(5+\frac{1}{4\omega^2}\right)+\frac{\lambda}{3}\left(\kappa^2\right)^2\Lambda\left(20\omega+15-\frac{1}{2\omega}\right)\,,
\nonumber\\
&&\gamma=\lambda\left(\frac{10}{3}\omega-\frac{13}{6}-\frac{1}{4\omega}\right)\,.\label{betabeta}
\end{eqnarray}
In general, $\kappa^2(t')\,,\lambda(t')\,,\Lambda(t')$ are assumed to be positive. On the other hand,  
$\omega(t')$ is expected to be negative to have a positive
contribution from $R^2$ (like in Starobinsky model).

The RG parameter $t'$ reads
\begin{equation}
t'=\frac{t'_0}{2}\log \left[\frac{R}{R_0}\right]^2\,,\quad 0<t_0'\,,
\label{tprime}
\end{equation}
where $t'_0$ is a positive dimensionless constant and $R_0$ is the mass
scale for the Ricci scalar.  We will set $R_0$ as the value of the Ricci scalar in
our dark energy universe, namely $R_0=4\Lambda$, $\Lambda$ being the Cosmological Constant: it means that $t'(R=R_0)=0$ today, while in the past
$0<t'(R_0<R)$. 

The first equation in~(\ref{couplingconst}) immediatly leads to
\begin{equation}
\lambda(t')=\frac{\lambda(0)}{1+\lambda(0)\beta_2 t'}\,,\label{lambda}
\end{equation}
where $\lambda(0)$ is an integration constant and $\lambda(t'=0)=\lambda(0)$.
Moreover, by analyzing the asymptotic behaviour 
of the implicitly-given effective coupling constants
$\omega(t')\,,\kappa^2(t'), \Lambda(t')$, when $t'\rightarrow\infty$, namely in
the high curvature limit of inflation\footnote{This result emerges from the analysis of the fixed points of $\omega(t')$ which obeys to the second equation in (\ref{couplingconst}). In particular, one finds that $\omega(1\ll t')\simeq-0.02+c_0/(1+\lambda(0)\beta_2 t')^{1.36}$, $c_0$ being a constant.}, one finds~\cite{due}
\begin{eqnarray}
\omega\simeq-0.02\,,\quad
\kappa^2(t')\simeq\kappa_0^2(1+\lambda(0)\beta_2 t')^{0.77}\,,\quad
\Lambda(t')\simeq\Lambda_0\frac{1}{(1+\lambda(0)\beta_2
t')^{0.55}}\,.\label{setuno}
\end{eqnarray}
Therefore, it is natural to fix the constants $\kappa^2_0$ and $\Lambda_0$ as
\begin{equation}
\kappa^2(t'=0)\equiv\kappa^2_0=\frac{16\pi}{M_{\text{Pl}}^2}\,,\quad
\Lambda(t'=0)\equiv\Lambda_0=2\Lambda\,,\label{setdue}
\end{equation}
where, as usually, $M_{\text{Pl}}$ is the Planck Mass\footnote{By using the notation of the preceding sections we also identify $\kappa_0\equiv\kappa$.}, and $\Lambda$ is the Cosmological Constant
of dark energy whose contribute can be neglected during inflation.\\
\\
Let us return to the flat FRW metric (\ref{metric}). Since the contribute of the Weyl tensor is null, we can treat our theory like $F(R)$-gravity, and we derive the gravitational equation
\begin{eqnarray}
0&=&
-\Lambda(t')+\frac{6 H^2}{\kappa^2(t')}-\frac{6
H}{(\kappa^2(t'))^2}\frac{d\kappa^2(t')}{d t'}\left(\frac{t_0' \dot
R}{R}\right)
+\frac{\omega(t')}{3\lambda(t')}
\left[6R\dot H
-12H\dot R\right]\nonumber\\&&\hspace{-2cm}
-12 H\frac{d}{d t'}\left(\frac{\omega(t')}{3\lambda(t')}\right)\left(\dot R
t_0'\right)
+6\left(H^2+\dot H\right)\Delta(t')\frac{t_0'}{R}-6 H
\left[\frac{d\Delta(t')}{d
t'}\left(\frac{t_0'}{R}\right)^2-\Delta(t')\frac{t_0'}{R^2}\right]\dot R\,,
\label{NN2}
\end{eqnarray}
where $t'$ is a function of $R$ as in (\ref{tprime}) and 
\begin{equation}
\Delta(t')=\left[\frac{R}{(\kappa^2(t'))^2}\frac{d\kappa^2(t')}{d
t'}+R^2\frac{d}{d
t'}\left(\frac{\omega(t')}{3\lambda(t')}\right)+\frac{d\Lambda(t')}{d
t'}\right]\,,\quad
R=12H^2+6\dot H\,.\label{Delta}
\end{equation}
Note that (\ref{NN2}) corresponds to the first equation in (\ref{EOMsgen}) with the first expression in (\ref{rhopMG}). As we know, in the absence of matter, such an equation is enough to describe the evolution of the whole system.\\
\\
During inflation $R_0\ll R$, namely $1\ll t'$ in (\ref{tprime}). If we assume that $R\simeq 12H_\text{dS}^2$ is almost a constant,
from (\ref{NN2})--(\ref{Delta}) 
with
(\ref{lambda})--(\ref{setdue}) and $\Lambda_0=0$, we derive the following de Sitter solution,
\begin{equation}
H_\text{dS}^2\kappa_0^2\simeq
\frac{0.0146}{t_0'(\lambda(0)t')^{0.77}}\,.\label{setH}
\end{equation}
For example, if we put $t_0'=\lambda(0)=1$ and we take $R\sim 10^{120} R_0$, we find
$H_\text{dS}^2\kappa_0^2\simeq 19\times
10^{-5}$, which leads to a curvature close to the Planck energy. 

As an interesting feature of the model, we find that such a solution is always unstable and early-time acceleration disappears after some times. 
If we perturb the Hubble parameter as
\begin{equation}
H=H_\text{dS}+\delta H(t)\,,\quad |\delta H(t)|\ll1,\label{approxH}
\end{equation}
and we assume
\begin{equation}
1\ll(H_\text{dS}\kappa_0)^2 t'^{2.27}\,,\label{condHtprime}
\end{equation}
Eq.~(\ref{NN2}) with
(\ref{lambda})--(\ref{setdue}), $\Lambda_0=0$ and $H_{\text{dS}}$ as in (\ref{setH}) leads at the first order in $\delta
H\equiv \delta H(t)$,
\begin{equation}
D_0\delta H+
t'[19.152 (H_\text{dS}\kappa_0)(\kappa_0\dot \delta
H)+6.384(\kappa_0^2\ddot\delta H)]\simeq 0\,,\label{stabeq}
\end{equation}
where
\begin{equation}
D_0=
      \left(\frac{0.223}{(\lambda(0) t')^{0.77}}-30.528 t_0'
\left(H_{\text{dS}}\kappa_0\right)^2\right)\,.\label{D0}
\end{equation}
The solution of this equation is given by
\begin{equation}
\delta H= h_\pm\exp\left[A_\pm t\right]\,,
\quad
A_\pm=
\left[\frac{H_{\text{dS}}}{2}\left(-3\pm\sqrt{9-\frac{0.627
D_0}{(H_\text{dS}\kappa_0)^2t'}}\right)\right]\,,
\quad
|h_{\pm}|\ll 1\,,\label{solpert}
\end{equation}
where $h_\pm$ are constants corresponding to  plus and minus
signs inside $A_\pm$.
Thus, the solution is unstable under
the condition
\begin{equation}
D_0<0\,.\label{D}
\end{equation}
Now we note that if
\begin{equation}
\frac{0.007}{t_0(\lambda(0) t')^{0.77}}<\left(H_\text{dS}\kappa_0\right)^2\,,
\end{equation}
both of the conditions (\ref{condHtprime}) and (\ref{D}) are satisfied: but by using (\ref{setH}), we can easily see that this formula holds always true
independendently on the bound of inflation from $\lambda(0)$! 
Moreover, one has
\begin{equation}
A_+\simeq 0.796\frac{H_\text{dS}t_0'}{t'}\,,\quad A_-\simeq -3
H_\text{dS}\,,\label{Apiu}
\end{equation}
where  $D_0$ has been considered very small. Inflation finishes after $\Delta t\simeq 1/A_+$. For example, if $t_0'=\lambda(0)=1$ and $R\sim 10^{120} R_0$, we obtain $\Delta t_\text{f}\simeq 18\times 10^4/M_{\text{Pl}}$. The $e$-folds number of inflation can be estimated as
\begin{equation}
\mathcal N\simeq H_\text{dS}\Delta t\simeq
\frac{H_\text{dS}}{A_+}
\simeq 1.26\left(\frac{t'}{t_0'}\right)\,,\label{NNNN}
\end{equation}
where $1\ll t'$ depends on the curvature. As a result, the $e$-folds number is extremelly large and, since the spectral index $n_s$ in (\ref{spectralMG}) for this kind of model of modified gravity behaves as $n_s\sim 1-2/\mathcal N$, the Planck data cannot be satisfied (see Ref.~\cite{due} for details).\\ 
\\
A possible solution of the problem is to consider in the action (\ref{actionQ}) the Gauss-Bonnet and the $\Box R$-terms which
have been omitted above and give contribution when their coefficients are not constant. We add the following piece to the action,
\begin{equation}
I_{G\,,\Box R}=-\int_\mathcal M d^4 x\sqrt{-g} \left[\gamma(t') G-\zeta(t')\Box
R\right]\,,\label{new}
\end{equation}
where $\gamma(t')\,,\zeta(t')$ are effective
coupling  constants.
In analogy with (\ref{setuno}) we may assume (see also the recent work in Ref.~\cite{GBdisc}),
\begin{equation}
\gamma(t')=\gamma_{0}(1+c_1
t')\,,\quad\zeta(t')=\zeta_0(1+c_2 t')\,,\label{gammazetalimit}
\end{equation}
with $\gamma_0\,,\zeta_0$ constants and $c_{1,2}$ numerical
coefficients.

Now Eq.~(\ref{NN2}) reads
\begin{eqnarray}
0&=&
-\Lambda(t')+\frac{6 H^2}{\kappa^2(t')}-\frac{6
H}{(\kappa^2(t'))^2}\frac{d\kappa^2(t')}{d t'}\left(\frac{t_0' \dot
R}{R}\right)
+\frac{\omega(t')}{3\lambda(t')}
\left[6R\dot H
-12H\dot R\right]
-12 H\frac{d}{d t'}\left(\frac{\omega(t')}{3\lambda(t')}\right)\left(\dot R
t_0'\right)\nonumber\\&&
+6\left(H^2+\dot H\right)\Delta(t')\frac{t_0'}{R}-6 H
\left[\frac{d\Delta(t')}{d
t'}\left(\frac{t_0'}{R}\right)^2-\Delta(t')\frac{t_0'}{R^2}\right]\dot R
-24 H^3\frac{d\gamma(t')}{d t'}\frac{t'_0\dot R}{R}-6
H\left[\frac{d\gamma(t')}{d t'}\frac{t_0'\dot G}{R}\right]\nonumber\\&&
-3\mathcal A\dot R^2
-2\mathcal B\dot R^2 R
+6\frac{d}{d t}\left[2\mathcal A\left(4H^2+3\dot H\right)\dot R+\mathcal B
H\dot R^2\right]
+18 H\left[2\mathcal A\left(4H^2+3\dot H\right)\dot R+\mathcal B H\dot
R^2\right]\nonumber\\
&&-36\left(3H^2+\dot H\right)\mathcal A\,H\dot R-72H\frac{d}{d t}\left(\mathcal
A H \dot R\right)-12\frac{d^2}{d t^2}
\left(\mathcal A H \dot R\right)
\,,
\label{NN2new}
\end{eqnarray}
where $\Delta (t')$ is still given by (\ref{Delta}) and
\begin{equation}
\mathcal A=\left(\frac{d\zeta(t')}{d t'}\frac{t'_0}{R}\right)\,,\quad
\mathcal B=\left[\frac{d^2\zeta(t')}{d
t'^2}\left(\frac{t'_0}{R}\right)^2-\frac{d\zeta(t')}{d
t'}\frac{t'_0}{R^2}\right]\,.\label{NN2new2}
\end{equation}
If we use
(\ref{NN2new})--(\ref{NN2new2}) with
(\ref{lambda})--(\ref{setdue}) and (\ref{gammazetalimit}), $\Lambda_0=0$ and $1\ll t'$,  we derive
the de Sitter
solution for inflation
\begin{equation}
H_\text{dS}^2\kappa_0^2\simeq
\frac{322.762}{\left(22085.2-34725.2(d\gamma(t')/dt')\right)t_0'(\lambda(0)t')^{0.77}}\,,\quad\frac{d\gamma(t')}{d
t'}<0\,,\label{dS2}
\end{equation}
where we have considered  $|d\gamma(t')/d t'|\ll t'^2$ with $d\gamma(t')/d t'<0$, namely
$\gamma_0c_1<0$ in (\ref{gammazetalimit}). Note that the $\Box R$-term does not contribute to the de Sitter
solution.
By perturbating the solution as in
(\ref{approxH}), we get
\begin{equation}
\tilde D_0\delta H+
t'[19.152 (H_\text{dS}\kappa_0)(\kappa_0\dot \delta
H)+6.384(\kappa_0^2\ddot\delta H)]\simeq 0\,,\label{stabeq2}
\end{equation}
where
\begin{equation}
\tilde D_0=
      \left[\frac{0.223}{(\lambda(0) t')^{0.77}}-(30.528-48 d\gamma(t')/d t')
t_0' \left(H_{\text{dS}}\kappa_0\right)^2\right]\,.\label{D02}
\end{equation}
The solution of such equation is given by
\begin{equation}
\delta H= h_\pm\exp\left[\tilde A_\pm t\right]\,,
\quad
\tilde A_\pm=
\left[\frac{H_{\text{dS}}}{2}\left(-3\pm\sqrt{9-\frac{0.627 \tilde
D_0}{(H_\text{dS}\kappa_0)^2t'}}\right)\right]\,,
\quad
|h_{\pm}|\ll 1\,.\label{solpert2}
\end{equation}
Also in this case, $h_\pm$ are the integration constants corresponding to the
signs plus and
minus inside $\tilde A_\pm$ and the solution is unstable if $\tilde D_0<0$. Again, it is easy to verify from (\ref{dS2}) that the instability condition is always satisfied 
independently on the value of $d\gamma(t')/(t')$, and the model can exit from the early-time acceleration. One also has
\begin{equation}
\tilde A_+\simeq 36019\times
10^{-9}\frac{H_\text{dS}t_0'}{t'}\left(22085.2-34725.2 d\gamma(t')/dt'\right)\,,\quad
\tilde A_-\simeq -3 H_\text{dS}\,,\label{Apiu}
\end{equation}
where  $\tilde D_0$ has been considered very small. Thus, thanks to the contribute of the Gauss-Bonnet, the $e$-folds in (\ref{NNNN}) decreases when
$\tilde A_+$ (i.e. $|d\gamma(t')/dt'|=|\gamma_0 c_1|$) increases, and one may finally get a correct value of the spectral index in agreement with the Planck result. Since the tensor-to-scalar ratio is typically very small ($r \sim1/\mathcal N^2$) and also satisfies the cosmological data, we conclude that our higher derivative gravitational theory with the account of quantum effects can be considered a good candidate for realistic inflation.

\section{Conclusions}

In this review we have revisited inflation in $F(R)$-gravity. A realistic inflationary scenario must lead to the perturbations at the origin of the anisotropy of our universe. In the specific, the spectral index and the tensor-to-scalar ratio at the end of inflation must be in agreement with the last Planck satellite observations. By starting from the values of this parameters and by assuming a reasonable Ansatz for the effective EoS parameter of the universe during the primordial acceleration, it is found that the $F(R)$-inflation has to be realized nearly to the Starobinsky model with the account of a quadratic correction in the Ricci scalar to the Einstein's theory: in the scalar field representation, it is analogous to an exponential potential respect to the field. A Starobinsky-like model with cosmological constant which possesses an exact solution for realistic inflation without making use of slow-roll approximation has been derived. 

In the second part of the work, we have investigated the possible modifications to Starobinsky inflation by working on its scalar field representation.
Chaotic inflation is realized by a constant in the potential of the Einstein frame which corresponds to the $R^2$-term in the Jordan frame. The model $F(R)\sim R^2$ possesses an exact de Sitter solution for an arbitrary value of the curvature, which is fixed by the initial conditions at the time of the Big Bang. In order to exit from inflation, we need an extra term which makes inflation unstable. In the Einstein frame, this role can be played by terms proportional to $\sim\exp\left[n\kappa\phi\right]$, $0<n<2$ in the potential, namely 
terms proportional to $R^{2-n}$, $0<n<2$ in the $F(R)$-framework. The case $n=1$ corresponds to the Starobinsky model. Different choices of $n$ do not change the spectral index of Starobinsky inflation, but affect the value of the tensor-to-scalar ratio: in order to satisfy the Planck data we have to take $1/5<n<2$ (see also the recent work in Ref.~\cite{Broy}). 
It is important to note that in the low curvature limit of the theory the scalar potentials under investigation correspond to Einstein's gravity in the frame of $F(R)$-gravity, even when $n\neq 1$.
Other works on inflation near to $R^2$-model can be found in Ref.~\cite{Zerg}.
The methods developed in this work maybe equally applied to
$F(R)$-gravity with Lagrange multiplier constraint~\cite{LLL}, which includes
mimetic $F(R)$-gravity for inflation as specific case~\cite{OdL}. An useful work where various theoretical
and observational consequences in $F(R)$-gravity are discussed in detail can be found in Ref.~\cite{addd}.

In the last part of the work, we have considered early-time acceleration in a theory of higher-derivative quantum gravity, where several terms quadratic in the curvature invariants have been taken into account (for $F(R)$-gravity in the presence of trace anomaly see also Ref.~\cite{miotrace}). As an interesting general feature of the model, we find that the theory possesses an unstable de Sitter solution. Moreover, thanks to the Gauss-Bonnet contribute, one may reproduce the Planck data and inflation results to be viable.

A recent work on $F(R)$-gravity and scalar field inflation can be found in Ref.~\cite{FphiR}. In Ref.~\cite{ROd} it was also proposed a model of singular inflation which maybe also
realized in $F(R)$-gravity.


\end{document}